\RequirePackage{fix-cm}
\documentclass[smallextended]{svjour3}   

\smartqed  % flush right qed marks, e.g. at end of proof

\usepackage{amssymb}
\usepackage{hyperref}
\usepackage{amsmath}
\usepackage{graphicx}
\usepackage[english]{babel}
\usepackage{siunitx}
\usepackage[labelfont=bf]{caption}
\providecommand{\keywords}[1]
{
	\small	
	\textbf{\textit{Keywords---}} #1
}

\usepackage[a4paper,includeheadfoot,margin=2.54cm]{geometry}

\title{Fiber Optical Shape Sensing of Flexible Instruments for Endovascular Navigation}
\author{Sonja~J\"ackle$^1$ \and 
Tim~Eixmann$^2$ \and 
Hinnerk~Schulz-Hildebrandt$^{2,3,4}$ \and  Gereon~H\"uttmann$^{2,3,4}$ 
\and 
Torben~P\"atz$^5$ }

\date{\today}

\institute{ $^1$Fraunhofer MEVIS, Institute for Digital Medicine, L\"ubeck, Maria-Goeppert-Stra\ss e 3, 23562 L\"ubeck, Germany \\
            $^2$Medical Laser Center L\"ubeck GmbH,  Peter-Monnik-Weg 4, 23562 L\"ubeck, Germany \\
            $^3$Institute of Biomedical Optics, Universit\"at zu L\"ubeck, Peter-Monnik-Weg 4, 23562 L\"ubeck, Germany \\
            $^4$German Center for Lung Research, DZL, Airways Research Center North, 22927 Gro\ss hansdorf, Germany \\            
            $^5$Fraunhofer MEVIS, Institute for Digital Medicine, Bremen, Am Fallturm 1, 28359 Bremen, Germany \\
}

\begin{document}
\maketitle

\begin{abstract}:\\
% ToDo: Abstract an Anforderungen von IJCARS anpassen:
%Please provide a structured abstract of 150 to 250 words which should be divided into the following sections:
%    Purpose (stating the main purposes and research question)
%    Methods
%    Results
%    Conclusions
%Please note that, contrary to the above text, the length of the Abstract should be about 200-300 word
\textit{Purpose:} \\
Endovascular aortic repair procedures are currently conducted with 2D fluoroscopy imaging. Tracking systems based on fiber Bragg gratings are an emerging technology for the navigation of minimal-invasive instruments which can reduce the x-ray exposure and the used contrast agent. Shape sensing of flexible structures is challenging and includes many calculations steps which are prone to different errors. To reduce this errors, we present an optimized shape sensing model. \\
\textit{Methods:} \\
We analyzed for every step of the shape sensing process, which errors can occur, how the error affects the shape and how it can be compensated or minimized. Experiments were done with a multicore fiber system with \SI{38}{cm} sensing length and the effects of different methods and parameters were analyzed. Furthermore we compared 3D shape reconstructions with the segmented shape of the corresponding CT scans of the fiber to evaluate the accuracy of our optimized shape sensing model. Finally we tested our model in a realistic endovascular scenario by using a 3D printed vessel system created from patient data. \\
\textit{Results:} \\ 
Depending on the complexity of the shape we reached an average error of \SIrange[range-units = single]{0.35}{1.15}{mm} and maximal error of \SIrange[range-units = single]{0.75}{7.53}{mm} over the whole \SI{38}{cm} sensing length. In the endovascular scenario we obtained an average and maximal error of \SI{1.13}{mm} and \SI{2.11}{mm}, respectively.  \\
\textit{Conclusions:} \\
The accuracies of the 3D shape sensing model are promising and we plan to combine the shape sensing based on fiber Bragg gratings with the position and orientation of an electromagnetic sensor system to obtain the located shape of the catheter.
\end{abstract}

\keywords{Fiber Bragg grating (FBG), shape sensing, flexible instruments, endovascular navigation.}

\section{Introduction:}
Cardiovascular diseases are the main cause of death in western industrial nations \cite{mendis2011global}. Some of these diseases like abdominal aortic aneurysms can be threaded by an endovascular aortic repair (EVAR) procedure, in which a stent is placed in the region of the aneurysm under 2D fluoroscopy. To reduce the X-ray exposure time and to supersede the angiography a three-dimensional navigation is needed.

Fiber Bragg grating (FBG) based systems are used for shape sensing, which enables three-dimensional navigation. FBGs are interference filters, which reflect a specific wavelength and are inscribed into the core of a optical fiber. Therefore, the change in reflected wavelength can be used to calculate strain. Combining multiple FBGs at the same longitudinal position allows to calculate curvature and direction angle. The most common configuration are three fibers arranged triangular around the structure to be measured \cite{henken2014,roesthuis2014}. This introduces significant errors due to possible changes in the core geometry \cite{henken2014}, which can be overcome by multicore fibers, where several cores are integrated into one fiber \cite{moore2012}. In addition other FBG types with different geometrical configurations have been introduced, as for example helically wrapped \cite{xu2016}. 

Most research groups use FBG systems for shape and force sensing of medical needles \cite{ourak2019}. These are short instruments, which have a simple bending profile allowing shapes with low bending and typically no torsion. For example Park~\cite{park2010} applied FBGs for shape sensing of biopsy needles and Roesthuis used it to reconstruct the shape of a nitinol needle~\cite{roesthuis2014}. A few works using optical fibers for flexible instruments have been reported in the literature: Shi~\cite{shi2014} used FBG systems together with EM-tracking and ultrasound to achieve a vasculature reconstruction and catheter modeling. Also Khan~\cite{khan2019multi} used four multicore fibers to reconstruct the first \SI{118}{mm} of a catheter. However, to our knowledge, there are currently no studies on the accuracy of fiber optical shape sensing for very long and flexible medical instruments.

In general shape reconstruction of flexible structures is more challenging, because higher deflections and torsion can occur. Thus the error analysis of the shape reconstruction from measured wavelengths to the reconstructed shape becomes more important. Also the accuracy of the measurement has to be very accurate, since the shape error accumulates along the instrument.

Therefore we introduce our optimized model for shape sensing of flexible instruments. Then we evaluated our model with 3D experiments. Finally, we tested it in a realistic endovascular scenario by inserting our fiber in a 3D printed aortic vessel system.

%%%%%%%%%%%%%%%%%%%%%%%%%%%%%%%%%%%%%%%%
%%-------------- METHODS -------------%%
%%%%%%%%%%%%%%%%%%%%%%%%%%%%%%%%%%%%%%%%

\section{Material and Methods}

\begin{figure}[ht]
	\centering
	\includegraphics[width=0.55\textwidth]{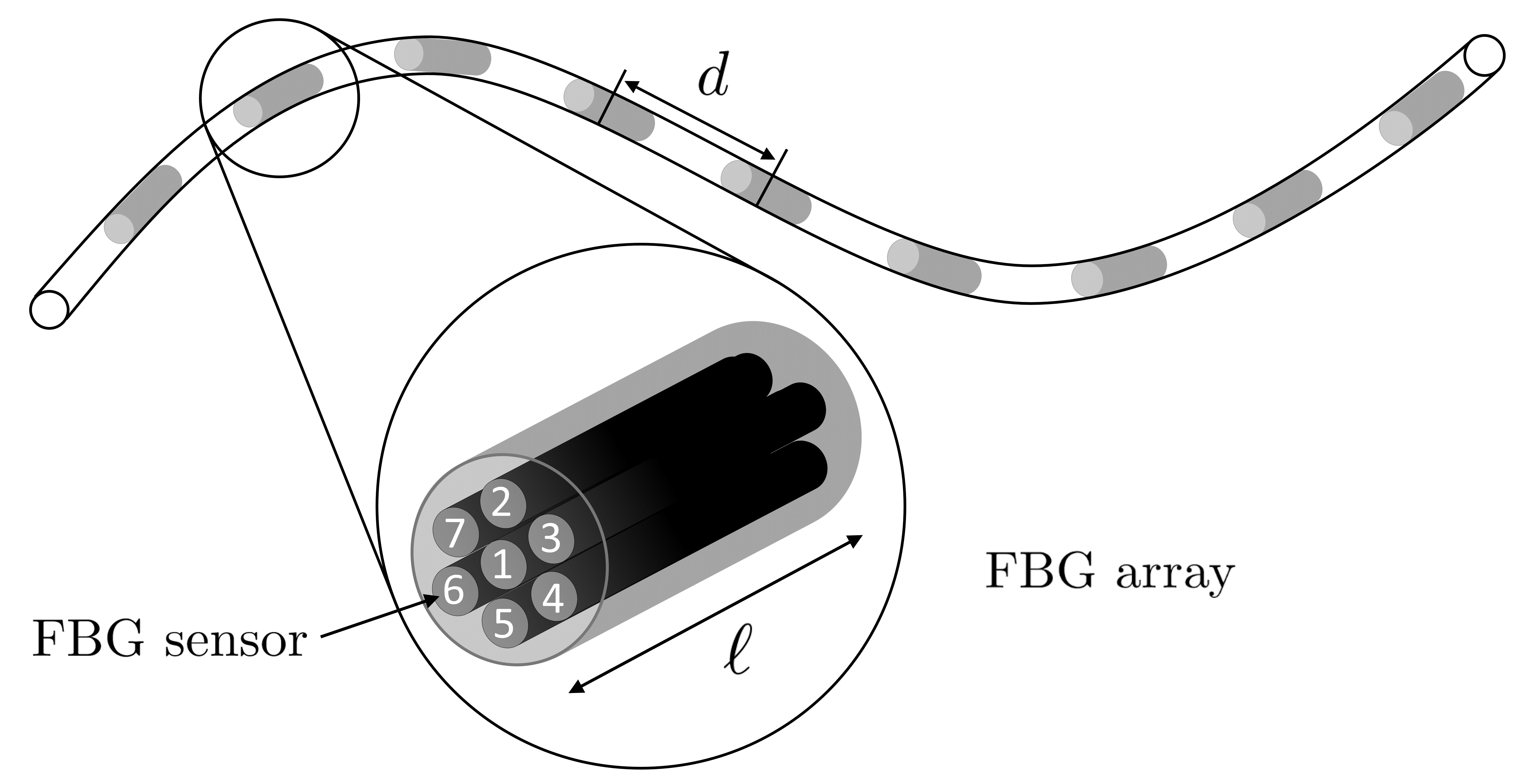}	
	\includegraphics[width=0.44\textwidth]{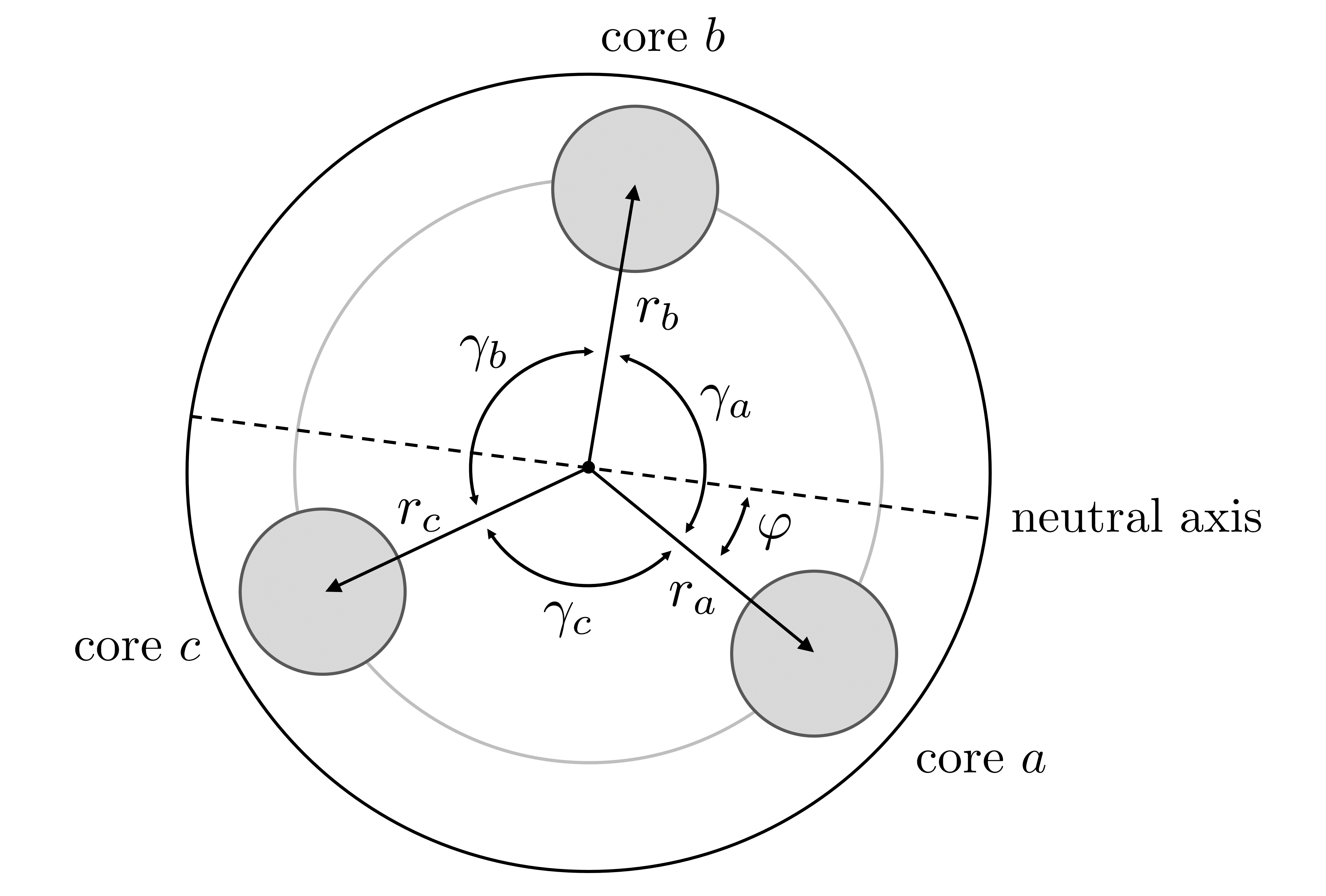}
	\caption{Left: A FBG system with center-to-center distance~$d$ and sensor length~$\ell$. The different cores are represented by numbers; Right: The cross section of a FBG system with triplet configuration}
	\label{fig-system}
\end{figure}
We consider a multicore fiber with $n$ FBG arrays along a flexible instrument, as shown in Fig.~\ref{fig-system} Each array contains seven FBGs, one center core and six outer cores. All FBGs have fixed length $\ell$ and the arrays are uniformly distributed with center-to-center distance $d$.

%%%%%%%%%%%%%%%%%%%%%%%%%%%%%%%%%%%%%%%%
%%--------------- MODEL --------------%%
%%%%%%%%%%%%%%%%%%%%%%%%%%%%%%%%%%%%%%%%
\subsection{Shape Sensing Model} 
We analyzed every shape sensing step and optimized it by minimizing the errors. The result is our optimized shape sensing model with the following steps:
\begin{enumerate}
	\item Wavelength shift calculation with reference wavelength.
	\item Strain computation for every core. 
	\item Strain interpolation for every core.
	\item Curvature and angle calculation by solving equations.
	\item Curvature and angle correction.
	\item Shape reconstruction with circle segments. 
\end{enumerate}
Every step is described in more detail in the following sections.

\subsubsection{Wavelength shift calculation}
FBGs are interference filters inscribed in short segments of the core of an optical fiber, which are able to reflect a specific wavelength of the incoming light \cite{lee2003}. The Bragg wavelength of a FBG is defined as
\begin{align*}
  \lambda_B = 2n_e \mathrm{\Lambda},
\end{align*}
where $n_e$ is the effective refractive index of the grating in the fiber core and $\mathrm{\Lambda}$ the grating period. Mechanical strain or temperature change the reflected wavelength, which is the basic idea for shape sensing. This results in a wavelength shift 
\begin{align*}
\mathrm{\Delta} \lambda = \lambda - \lambda_B
\end{align*}
of the current measured wavelength $\lambda$ in comparison to the reference wavelength $\lambda_B$ of the FBG. If the reference wavelengths of the FBGs are unknown, they have to be determined by a separate measurement where no strain is applied to the fiber system.

\subsubsection{Strain computation}
The measured wavelength shift $\mathrm{\Delta} \lambda_B$, which can be caused by an applied strain $\varepsilon$ or by a temperature change $\mathrm{\Delta} T$ in the Bragg gratings, is given by 
\begin{align*}
  \mathrm{\Delta} \lambda = \lambda_B \big((1-p_e)\varepsilon +(\alpha_{\mathrm{\Lambda}}+ \alpha_n)\mathrm{\Delta} T\big),
\end{align*}
where $p_e$ is the photo-elastic coefficient and $\alpha_{\mathrm{\Lambda}}$ and $\alpha_n$ are the thermal expansion coefficient and the thermo-optic coefficient of the fiber system \cite{hill1997}. Assuming a constant temperature $\mathrm{\Delta} T = 0$ simplifies the formula and allows to calculate the mechanical strain of the fiber with the measured wavelength shift:
\begin{align}
\mathrm{\Delta} \lambda_b = \lambda_b (1-p_e)\varepsilon. \label{shift}
\end{align}
The photo-elastic coefficient $p_e$ is directly related to the gauge factor $GF = 1 - p_e$. Photoelasticity is defined as the change in reflected wavelength depending on the mechanical strain applied in axial direction. For FBG systems the photo-elastic coefficient $p_e \approx 0.22$ can be found in the literature \cite{werneck2013}. Additionally, experiments have been described for determining the photo-elastic coefficient of any FBG system \cite{black2008}.

\subsubsection{Interpolation}
\label{sec:interpolation}
When curvature and angle direction are calculated for every FBG array, the required intermediate values can be determined by interpolation. This method assumes that the determined values of one FBG array are the values for one specific position, usually the center of the array.

Henken~\cite{henken2014} compared common interpolation methods for shape sensing and concluded that cubic spline interpolation is the best solution, which is currently the state-of-the-art interpolation. Interpolating the curvature is straight forward since it is continuous for any shape, whereas the direction angle interpolation is challenging for flexible structures, which may have discontinuous direction angle. 

Thus, we suggest to interpolate the strain, since it is continuous. Also, we use the averaged cubic interpolation, as introduced in \cite{jackle2019shape}: This yields a realistic interpolation based on the spatial properties of the FBGs by modeling the measured sensor values as an average over the sensor range. 

\subsubsection{Curvature and angle computation}
The calculation of the curvature and direction angle depends on the fiber system. The most common one is a triplet configuration \cite{henken2014,roesthuis2014}: Here the FBG system has three fiber cores with specific angles (typically \SI{120}{\degree}) in between, as illustrated in Fig.~\ref{fig-system}. 

For this configuration the relation between the strain and the curvatures and direction angles is described by the following equation system:
\begin{align*}
    \varepsilon_a &= - \kappa r_a \sin(\varphi)  + \varepsilon_0 \\
    \varepsilon_b &= - \kappa r_b \sin(\varphi + \gamma_a)  + \varepsilon_0 \\
    \varepsilon_c &= - \kappa r_c \sin(\varphi + \gamma_a + \gamma_b)  + \varepsilon_0,
\end{align*}
where $\varepsilon_x$ is the strain, $r_x$ the radius and $\gamma_x$ the angle of the corresponding fiber $x$. By solving the equation system we obtain the strain bias $\varepsilon_0$, the curvature $\kappa$ and the direction angle $\varphi$. The equation system can also be extended for four or more fibers.

The equation shows, that the curvature is influenced by the radii $r_x$ in a similar way as by the photo-elastic coefficient. The strain bias $\varepsilon_0$ includes a couple of effects: For the strain calculation we assumed a constant temperature, but conducting a measurement with another temperature than in the reference wavelength measurement, results in a bias. Also axial strain and pressure are part of the strain bias.

\subsubsection{Curvature and direction angle correction}
The determined curvatures and direction angles are influenced by various variables. To get the right values we suggest the following parameters for correction:
The curvature values are scaled by the photo-elastic coefficient $p_e$ and the center-to-core distances $r_x$. Since both parameters can be biased, we determine an correction parameter $c$ to get the right curvature values 
\begin{equation}
    \kappa_{\text{real}} = c \cdot \kappa.
    \label{eq-02}
\end{equation}
Also the fiber can be twisted during production or storage. But these twists are not contained in $\varepsilon_0$. Thus we obtain a measured direction angle 
\begin{equation}
    \varphi = \varphi_\text{real} + \varphi_\text{twist},
    \label{eq-03}
\end{equation}
which does not equal the real angle $\varphi_\text{real}$ because it is distorted by the twist angle $\varphi_\text{twist}$. Since it is an offset of the real angle $\varphi_\text{real}$ for every fiber it cannot be determined for FBGs in this geometrical configuration without a measurement, where $\kappa \neq 0$. Helically wrapped fibers include torsion in their model and the twist error can be compensated. For short and stiff instrument, this error is negligibly, whereas for flexible instruments the twist angles must be determined.

\subsubsection{Shape Reconstruction}
In the last years three different algorithms have been proposed for shape reconstruction: Moore~\cite{moore2012} presented a method based on the fundamental theorem of curves, which states that the shape of any regular three-dimensional curve, which has non-zero curvature, can be determined by its curvature and torsion \cite{banchoff2010}. It should be noted that the torsion of curves in mathematical contexts corresponds to the change of direction angle. The shape is obtained by solving the Frenet-Serret equations:
\begin{align*}
  \frac{dT}{dt} = \kappa N, \; \frac{dN}{dt} = - \kappa N + \tau B, \; \frac{dB}{dt} = -\tau N,
\end{align*}
where $\kappa$ is the given curvature, $\tau$ the given torsion, $T$ the tangent vector, $N$ the normal vector and $B$ the binormal vector of the given curve at the length position $t$. The integration of the determined tangent vectors yields the shape of the curve.
This method fails at points with $\kappa = 0$ and consequently the direction angle is undefined. Thus, this algorithm is not suitable for shape sensing of flexible structures.

Cui \cite{cui2018} suggested a method based on a parallel transport approach to overcome this. The equations to be solved are:
\begin{align*}
\frac{dT}{dt} = \kappa_1 N_1 + \kappa_2 N_2, \; \frac{dN_1}{dt} = - \kappa_1 T, \; \frac{dN_2}{dt} = -\kappa_2 T.
\end{align*}
where $\kappa_1$ and $\kappa_2$ are the curvature components corresponding to the normal vectors $N_1$ and $N_2$, which are orthogonal to the tangent vector $T$. The shape reconstruction is conducted in the same way as with Frenet-Serret.

Roesthuis~\cite{roesthuis2014} proposed another method based on circle segments: The shape is reconstructed by approximating it with elements of constant curvature. So for every element a circle segment of curvature $\kappa$ and length $l$ is created. Afterwards this segment will be rotated by the direction angle $\varphi$. By repeating this procedure for every given set $(\kappa, \varphi)$ we obtain the whole shape.

%%%%%%%%%%%%%%%%%%%%%%%%%%%%%%%%%%%%%%%%
%%------------ EXPERIMENTS -----------%%
%%%%%%%%%%%%%%%%%%%%%%%%%%%%%%%%%%%%%%%%

\subsection{Experimental methods}

For all experiments described below we used a multicore fiber system (FBGS Technologies GmbH) consisting of 7 cores, one center core and six outer cores each with an angle of $60$ degree in between, as shown in Fig.~\ref{fig-system}. It has 38 FBG arrays each with \SI{5}{mm} length and \SI{10}{mm} center-to-center distance, which are chains of draw tower gratings (DTG$\circledR$). 

In the next sections we used the following parameters and algorithms if they are not analyzed or specified there: We fixed our covered fiber to a precise ruler and used the measured wavelength as reference wavelengths, we used a photo-elastic coefficient $p_e = 0.22$ for strain calculation, we made averaged cubic interpolation of the strain values, we used four outer cores of our FBG system and we reconstructed the shape with circle segments. 

For matching reconstructed and ground truth shape we used the iterative closest point algorithm \cite{rusinkiewicz2001}. For evaluation we calculated the average and the maximum error defined as
\begin{align*}
e_{\text{avg}} := \frac{1}{n}\sum_{i=0}^{n} \| x_i - x^{\text{gt}}_i \|_2 
\text{ and }
e_{\text{max}} := \max ( \| x_0 - x_0^{\text{gt}} \|_2, \dots, \| x_n - x_n^{\text{gt}} \|_2),
\end{align*}
where $x_0, \dots, x_n$ are the reconstructed points and  $x_0^{gt}, \dots, x_n^{gt}$ are the measured ground truth points located every $10 \: \text{mm}$ along the shape.

\subsubsection{Wavelength shift computation}
For our multicore fiber we had no reference Bragg wavelengths given. Thus, we had to determine these wavelengths with a measurement without any strain. Therefore we analyzed the effect of the Bragg wavelength estimation: At different times we fixed the fiber in a straight line, measured the wavelengths, used it as reference Bragg wavelengths and reconstructed various types of shapes.

\subsubsection{Strain calculation}
The photo-elastic coefficient influences the shape by scaling the curvature. To analyze the effects of this parameter, we bent our fiber to varying degrees and reconstructed the shape using different $p_e$ values. 

\subsubsection{Interpolation}
For interpolation evaluation we formed our fiber to a snakelike shape, which has a few singularity points. Then we interpolated the measured strains as proposed in section \ref{sec:interpolation} and compared the resulting curvature and direction angles with the common interpolation methods.

\subsubsection{Curvature and angle computation}
Since we have a multicore fiber with 6 outer cores and one center core and an interrogator, where we can connect 4 cores, we do not have to use a triplet configuration with $120$ degrees in between. Thus we analyzed the effect of different combinations of 3 or 4 outer cores on the resulting curvatures and direction angles. 

\subsubsection{Curvature and angle correction}
To determine the twist angle $\varphi_{\text{twist}}$ we bent our fiber to 2D-shapes, where every position should have the same angle, as for example a bow shape, determined the direction angles and used it as twist angles, as described in Equation \eqref{eq-03}. To get the curvature scale factor $c$ we made several bow shapes with different radii, determined the best value assuming a photo-elastic coefficient $p_e = 0.22$ and used it for curvature correction, as described in Equation \eqref{eq-02}. 

\subsubsection{Shape reconstruction}
The shape reconstruction quality depends completely on the measured curvature and direction angles. When the measured values are correct, the proposed algorithms are able to reconstructed the correct corresponding shape. Therefore we analyzed the following two aspects:

First we looked at the convergence, i. e. how fine the segments in each iterative step of the algorithms must be to obtain the correct shape. Second we analyzed the noise handling of the three algorithms, i. e. how the results of the algorithms change with increasing gaussian noise. 
In both cases we simulated an arc shape with torsion, calculated the average curvature and median direction angle for every segment and reconstructed the shape. 

\subsubsection{3D shape reconstruction accuracy}
To evaluate our optimized shape sensing procedure we recorded 3D measurements: we covered our optical fiber (diameter: \SI{200}{\um}) with a metallic capillary tube (inner diameter: \SI{300}{\um} and total diameter: \SI{400}{\um}), fixed it in a specific shape, reconstructed the shape and compared it with the segmented ground truth from the CT images. For the endovascular experiment we inserted the FBG system into a 3D vessel system, which was created from a CT scan from a real patient. 

%%%%%%%%%%%%%%%%%%%%%%%%%%%%%%%%%%%%%%%%
%%-------------- RESULTS -------------%%
%%%%%%%%%%%%%%%%%%%%%%%%%%%%%%%%%%%%%%%%

\section{Results}
\subsection{Wavelength shift calculation}
Tab.~\ref{tab:references} shows the results of Bragg wavelength study. Here we reconstructed a straight line, a bended curve and a s-curve using different reference wavelengths. These wavelengths were determined experimentally by placing the fiber as straight as possible. For all three form types, the obtained errors of the reconstructions differ with the respective reference wavelengths used. This indicates that the measurement of the reference wavelengths for the calculation of the wavelength shift has to be very accurate. 
\begin{table}[h]
	\centering
	\begin{tabular}{lcccc}
	    \hline 
		Shape               & Error         & First Reference & Second Reference    & Third Reference     \\ 
		\hline
		Straight Line		& $e_{\text{avg}}$ &$0.36$           &$0.16$               & $2.28$      \\ 	
							& $e_{\text{max}}$ &$1.00$           &$0.30$               & $5.49$      \\ 
		Bended Curve		& $e_{\text{avg}}$ &$1.70$           &$1.56$               & $1.59$      \\    
		                    & $e_{\text{max}}$ &$4.92$           &$4.53$               & $4.82$      \\
	    S-Curve     		& $e_{\text{avg}}$ &$1.80$           &$1.76$               & $1.38$      \\    
		                    & $e_{\text{max}}$ &$4.58$           &$4.62$               & $3.06$      \\
		\hline 
	\end{tabular}
	\caption{Results of the Bragg wavelength study: Measured error $e_{\text{avg}}$ and $e_{\text{max}}$ in \SI{}{mm} for different shapes using various Bragg wavelengths}
	\label{tab:references}
\end{table} 

\subsection{Strain computation}
\begin{figure}[ht]
	\centering
	\begin{minipage}[t]{0.49\textwidth}
		\includegraphics[width=\textwidth]{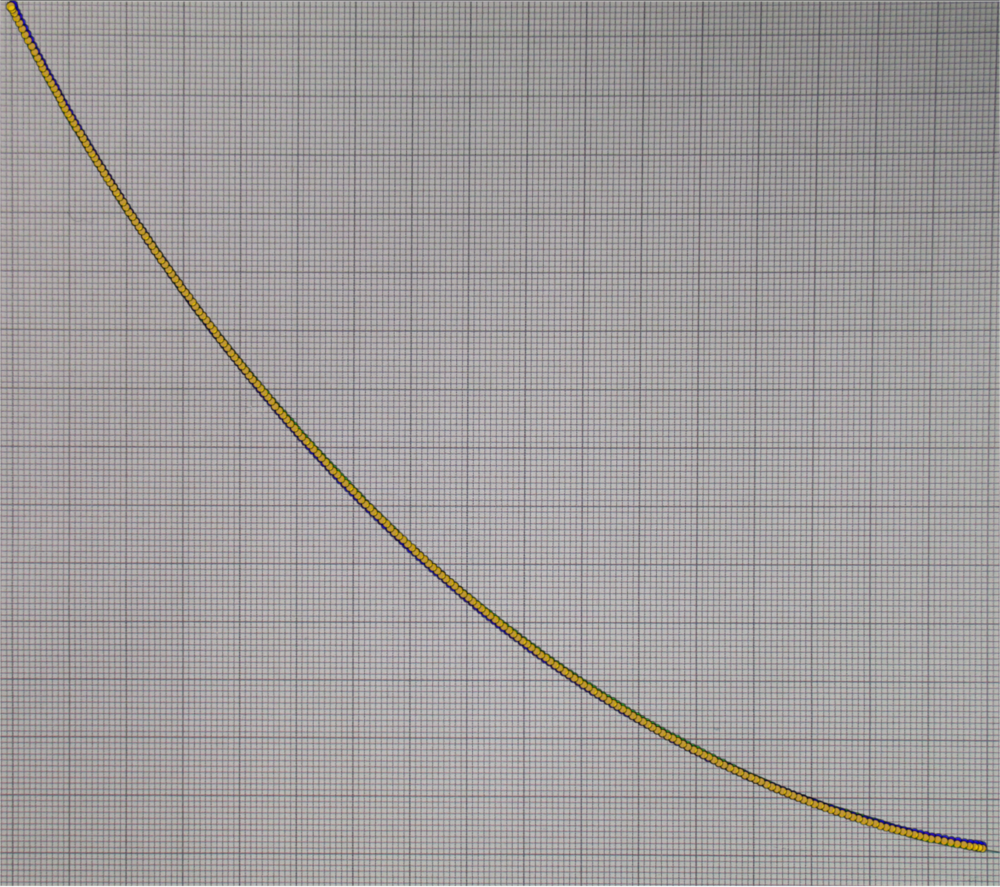}
	\end{minipage}
	\hfill
	\begin{minipage}[t]{0.49\textwidth}
    	\includegraphics[width=\textwidth]{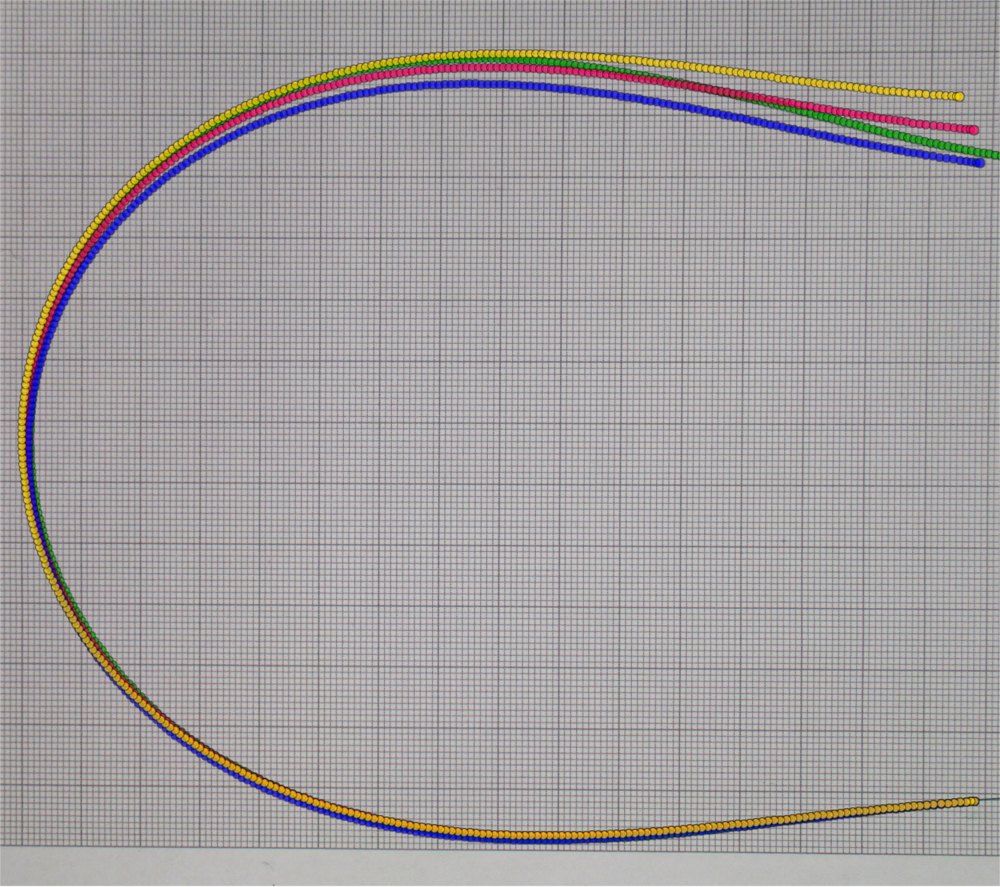}
	\end{minipage}
	\caption{Effect of the photo-elastic coefficient for different bending strengths (green = ground truth, yellow = reconstruction with $p_e = 0.21$, red = reconstruction with $p_e = 0.22$, blue = reconstruction with $p_e = 0.23$): The images show the fiber with low (first image) and high (second image) bending and the inserted projections of the reconstructed shape}
	\label{fig-photoelastic}
\end{figure}

The effect of the photo-elastic coefficient is shown in Fig.~\ref{fig-photoelastic} for shapes with different bending strengths. In the left image the reconstructed shape does not change notably, whereas in the right image the shape reconstructions differ significantly. Thus, the effect of the photo-elastic coefficient is higher in forms with high curvatures, while it has a minor effect in slightly curved structures. Thus this error has to be compensated.

\subsection{Interpolation}
\begin{figure}[ht]
	\centering
	\includegraphics[width=0.9\textwidth]{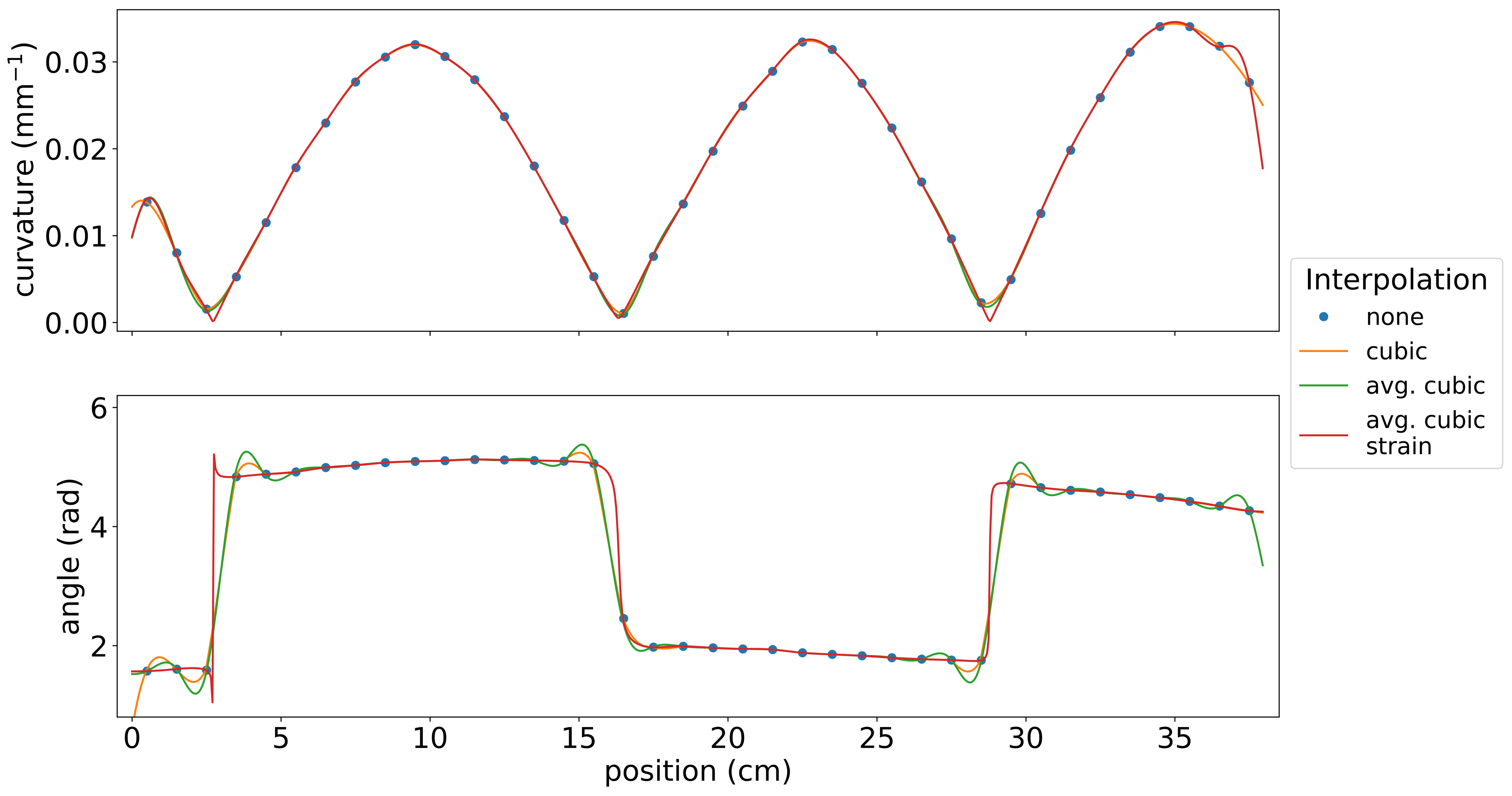}
	\caption{Results of interpolation study: the images show the resulting curvatures and direction angles of different interpolations methods along the fiber}
	\label{fig-interpolation}
\end{figure}

To evaluate the interpolation effect we interpolated the measured values of a snakelike shape, which has discontinuous direction angles. We compared our proposed method, calculating the strain with average cubic interpolation, with the state-of-the-art curvature and angle interpolation. The resulting curvatures and angles are shown in Fig.~\ref{fig-interpolation}. Interpolating the strain instead of curvature and torsion leads to more accurate interpolation: At discontinuity points the curvature is closer to zero and the direction angles are more accurate, whereas interpolating the direction angle results in overshoots before and after the discontinuity. 

\subsection{Curvature and angle computation}
We reconstructed the shape using three fiber cores with an angle of \SI{60}{degree} in between (cores 234), three fiber cores with an angle of \SI{60}{degree} and \SI{180}{degree} in between (cores 347) and four fiber cores with an angle of \SI{60}{degree} and \SI{180}{degree} in between (cores 2347). The numbers correspond to the cores as shown in Fig.~\ref{fig-system}. We observed, that configuration (347) leads not to sufficient results due to the linear dependency of two cores. For proper 3D reconstruction at least 3 linear independent core are needed as observed with configuration (234) and (2347). Using more cores leads to more stable results. Due to the low signal for shapes with high local bending radii, it is recommended to use more than three cores to ensure functionality for all possible shapes.

\subsection{Curvature and angle correction}
\begin{figure}[ht]
	\centering
	\includegraphics[width=0.47\textwidth]{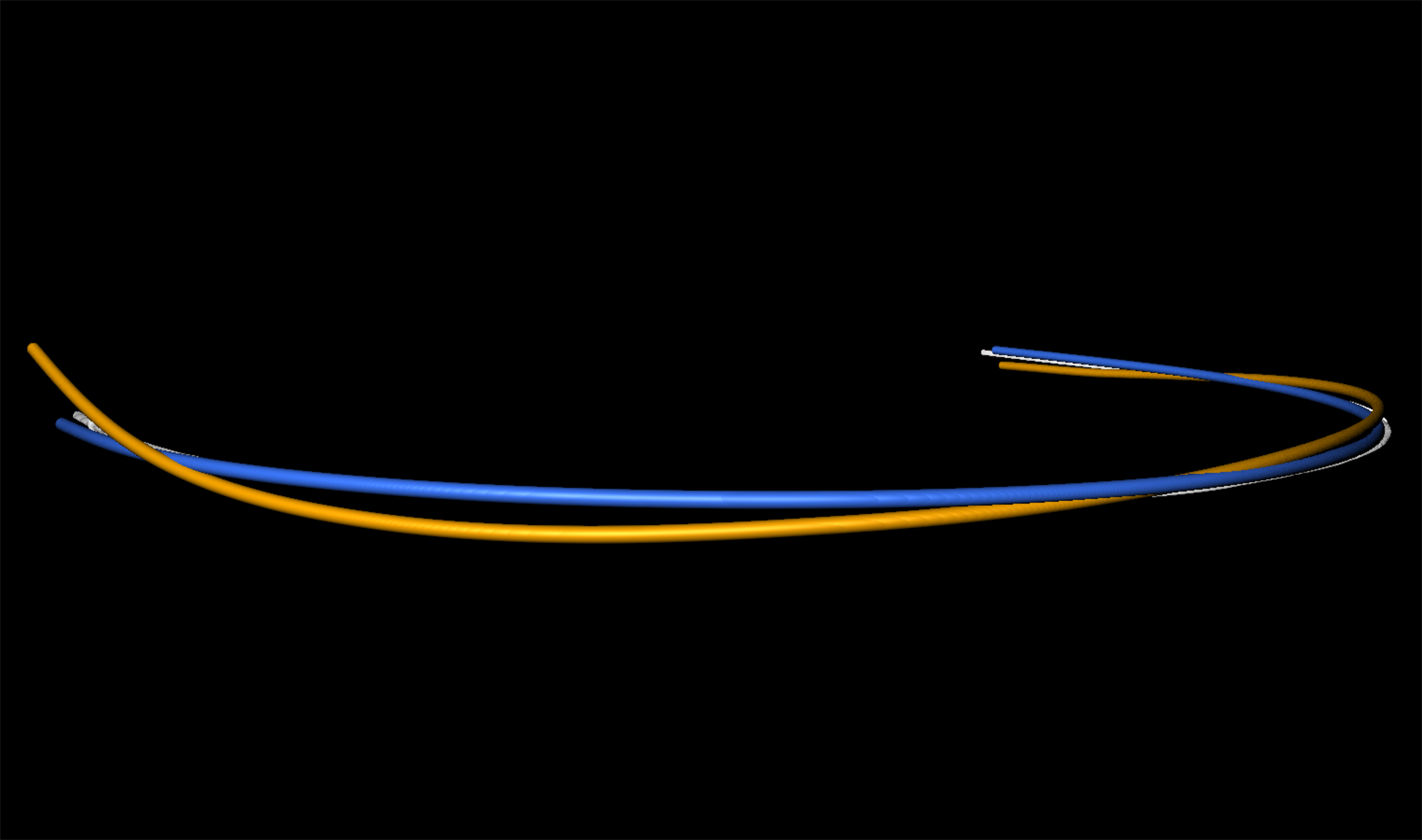}
	\includegraphics[width=0.52\textwidth]{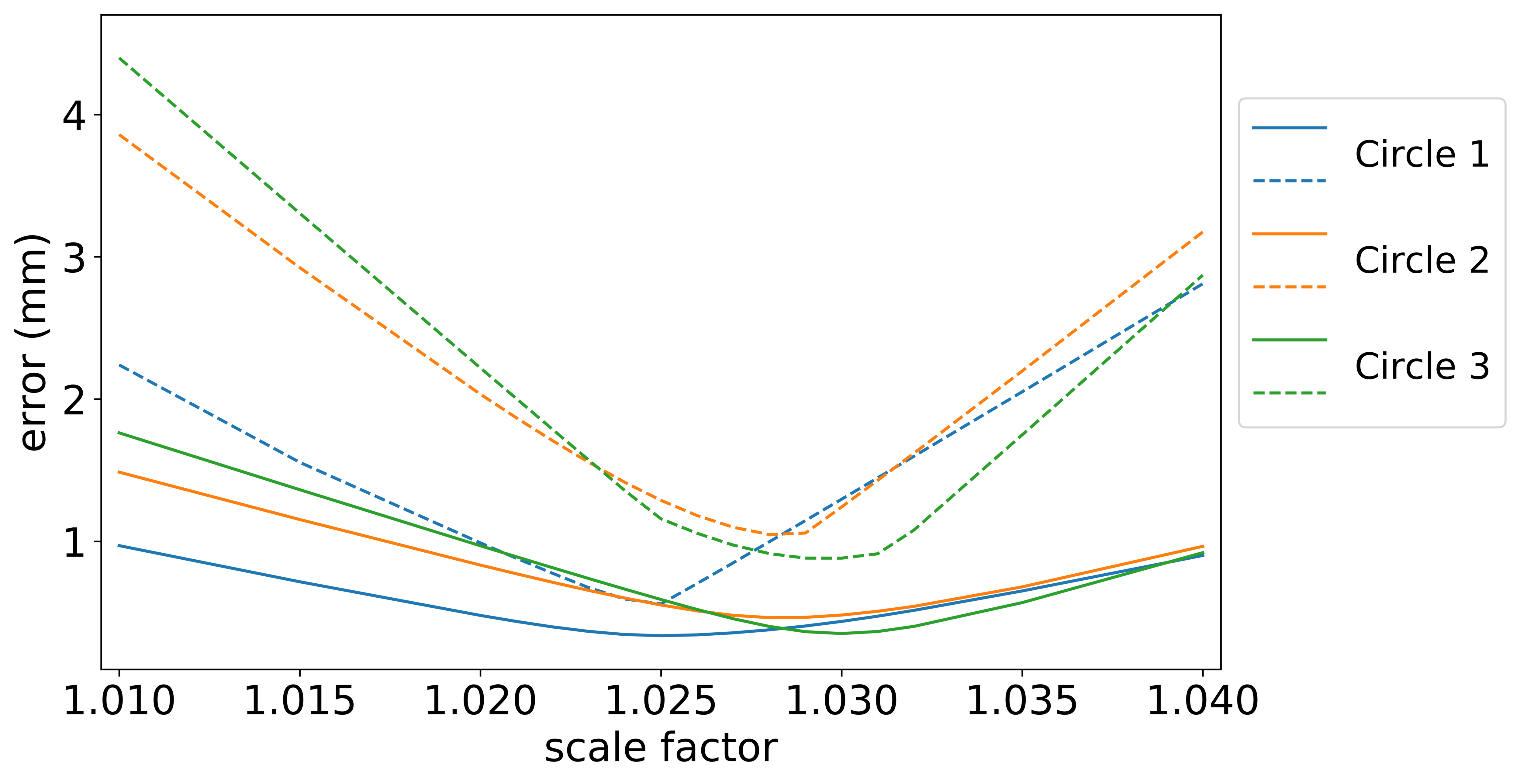}
	\caption{Left: Results of the twist angle study: The reconstructed shape with twist correction (blue) and without (orange) is shown. Ground truth from the CT scan is displayed in white; Right: Results of the curvature scale study: The average error (straight line) and maximum error (dashed line) are plotted for three different circles}
	\label{fig-curvatureScale}
\end{figure}

First we bent the fiber into a circular shape and used the determined angle as twist angle, as described in Equation \eqref{eq-03}. The result of this experiment is displayed in the left image of Fig.~\ref{fig-curvatureScale}: the reconstruction without angle correction is twisted whereas the corrected shape lies in the plane of the ground truth. Afterwards we made several circular shapes with various radii to determine the curvature scale factor of our FBG system. The results are shown in the right part of Fig.~\ref{fig-curvatureScale}. 
We found that a scale factor of $\approx 1.026$ achieves the best results and used it for curvature correction, as described in Equation \eqref{eq-02}.

\subsection{Shape reconstruction}
\begin{figure}[ht]
	\centering
	\includegraphics[width=0.45\textwidth]{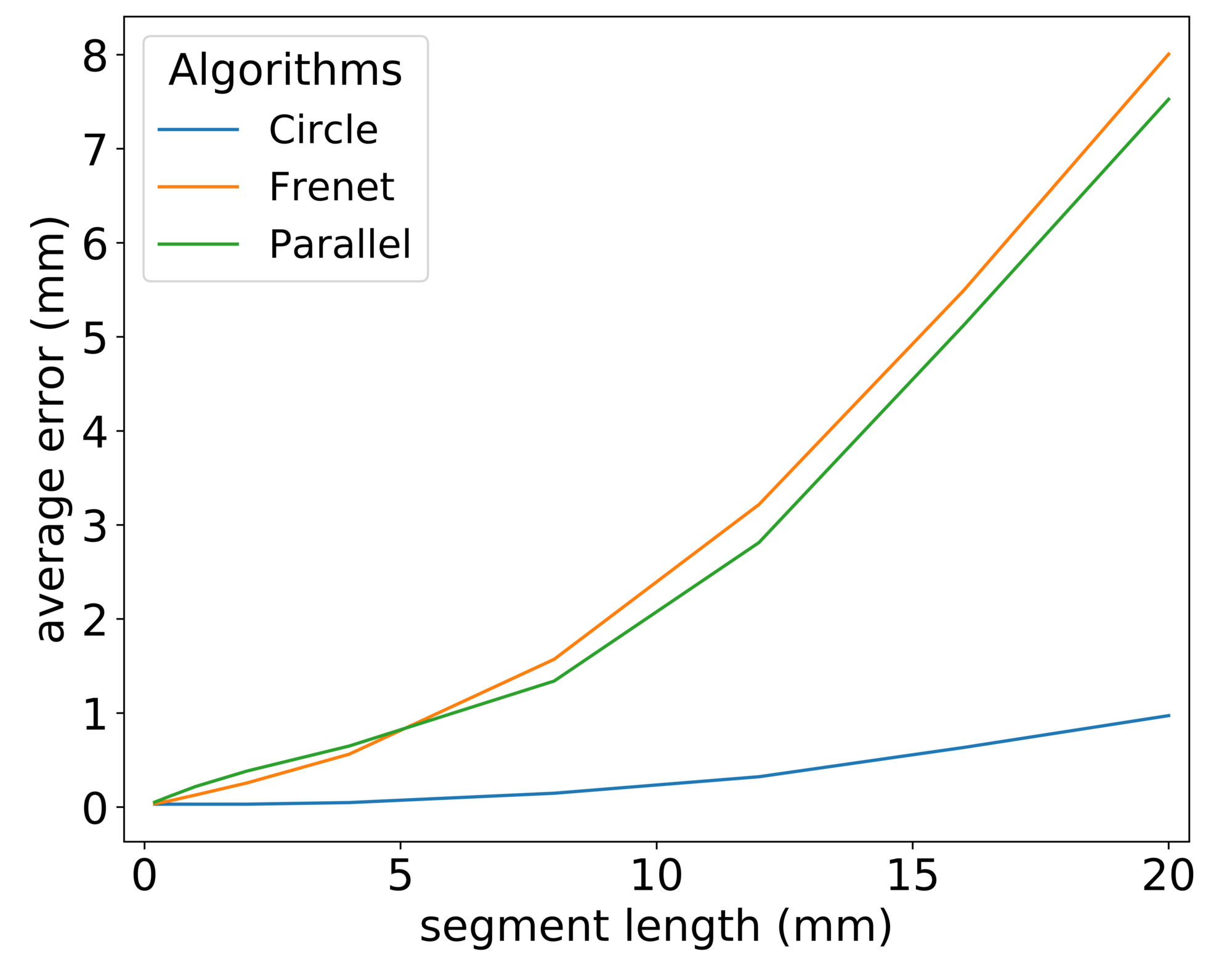}
    \includegraphics[width=0.45\textwidth]{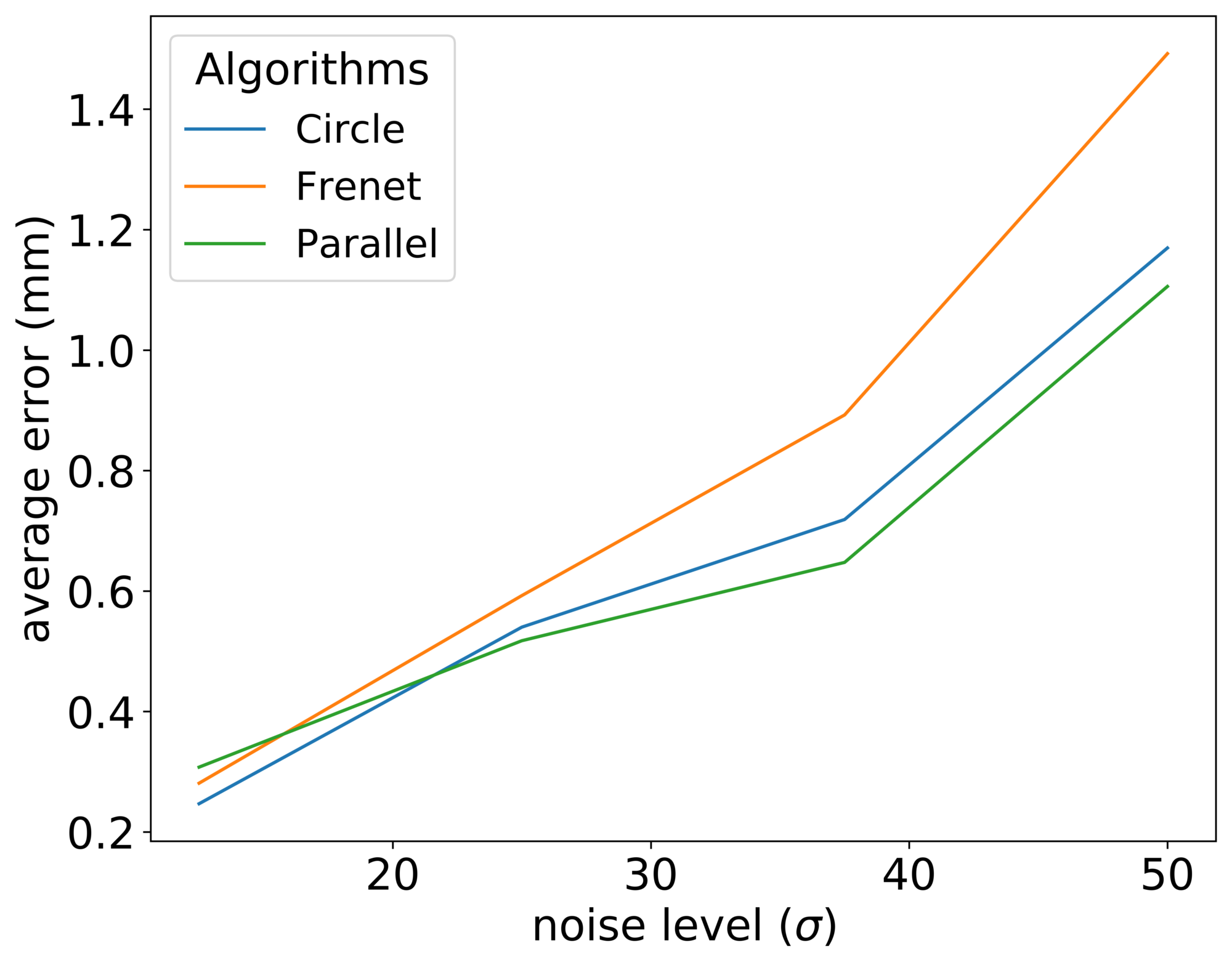}
	\caption{Shape reconstruction study results: The images show the average error as a function of segment length and of noise}
	\label{fig-reconstruction}
\end{figure}

The results of the convergence study, summarized in the left image of Fig.~\ref{fig-reconstruction}, show that the average error of all three algorithms increases with the segment length. Furthermore the numeric method based on circle segments has a faster convergence than Frenet-Serret and Parallel Transport. In the right image of Fig.~\ref{fig-reconstruction} the results of the noise study are are shown. Here we observe an increasing average error of all three algorithms with increasing noise. The methods based on Parallel Transport and circle segments have significantly better noise handling than Frenet-Serret. Thus we used the circle segment approach as shape reconstruction method.

\subsection{3D shape reconstruction accuracy}

\begin{figure}[ht]
	\centering
	\includegraphics[width=0.9\textwidth]{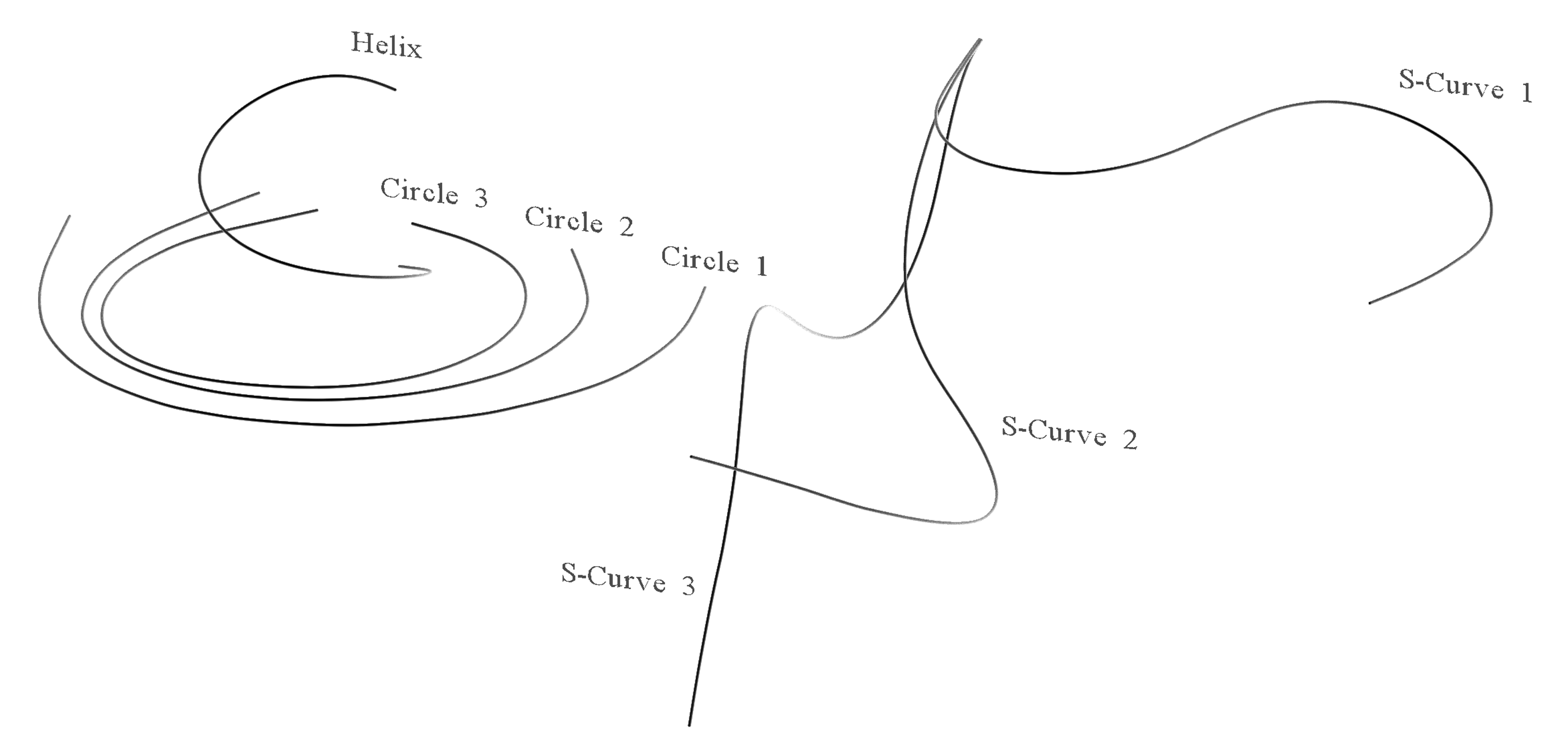}
	\caption{3D experiment with the fiber: the segmented shapes from the CT scan of the circle, s-curve and helix measurements are shown}
	\label{fig-3dshapes}
\end{figure}

For the 3D shape experiments we integrated the results of the previous experiments in our model.
We made several measurements bending our optical fiber to different 3D shapes. The segmented shapes from the CT scan of the circle, s-curve and helix measurements, as shown in Fig.~\ref{fig-3dshapes} were used as ground truth. The accuracies, shown in Tab.~\ref{tab:3dshapes}, depend on the complexity of the forms: For the circular shapes we obtain average error ${e_{\text{avg}} \approx \SI{0.5}{mm}}$ and maximal error $e_{\text{max}} \approx$ \SI{1}{mm}, whereas for s-curved and helical shapes we get higher errors, especially for s-curve 3 and helix.

Comparing our results with Khan~\cite{khan2019multi} we obtained higher errors. But Khan evaluated a catheter of only \SI{114}{mm} length in different configurations with weakly bending, constant or linear varying curvature and low or no torsion. Also they tested the catheter in no configuration with singularity points like s-curves. Hence the results of our 3D experiments with high deflections using the \SI{380}{mm} multicore fiber are nevertheless accurate and promising.

\begin{table}[htbp]
	\centering
	\begin{tabular}{lcc}
	    \hline
		Shapes $\setminus$ Errors (mm)   & $e_{\text{avg}}$ & $e_{\text{max}}$   \\ 
		\hline
		circle 1                         &           $0.35$ &         $0.75$     \\
		circle 2                         &           $0.50$ &         $1.15$     \\
		circle 3                         &           $0.50$ &         $1.02$     \\
		s-curve 1                        &           $0.70$ &         $1.29$     \\
		s-curve 2                        &           $0.57$ &         $1.98$     \\
		s-curve 3                        &           $1.15$ &         $7.53$     \\
		helix                            &           $1.00$ &         $4.72$     \\
		inside the vessel                &           $1.13$ &         $2.11$     \\
		\hline
	\end{tabular}
	\caption{Results of the 3D experiment: Measured errors $e_{\text{avg}}$ and $e_{\text{max}}$ in \SI{}{mm} for different 3D shape measurements}
	\label{tab:3dshapes}
\end{table} 

\begin{figure}[ht]	
	\centering
	\begin{minipage}[t]{0.497\textwidth}
		\centering
		\includegraphics[width=\linewidth]{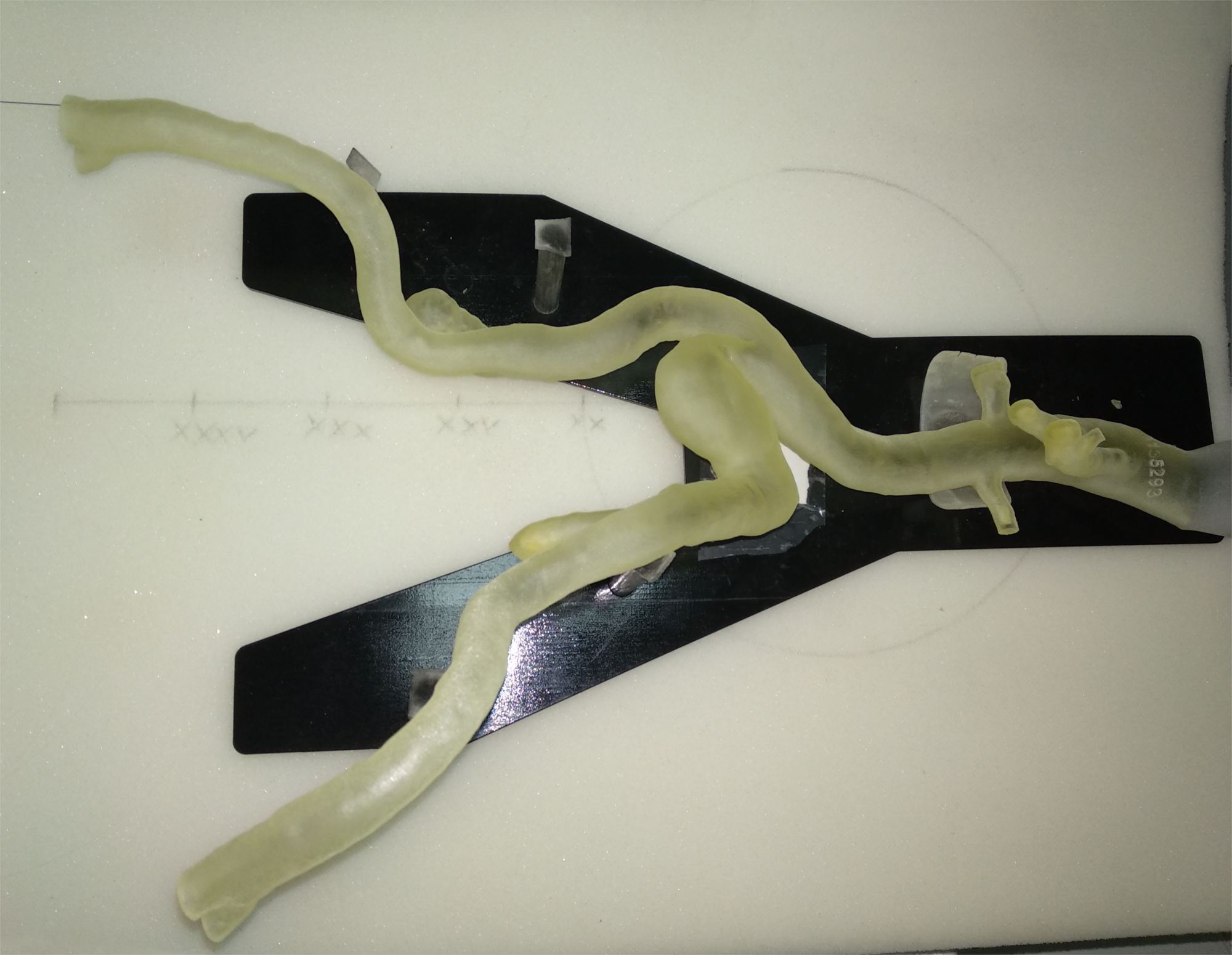}
	\end{minipage}\hfill
	\begin{minipage}[t]{0.49\textwidth}
		\centering
		\includegraphics[width=\linewidth]{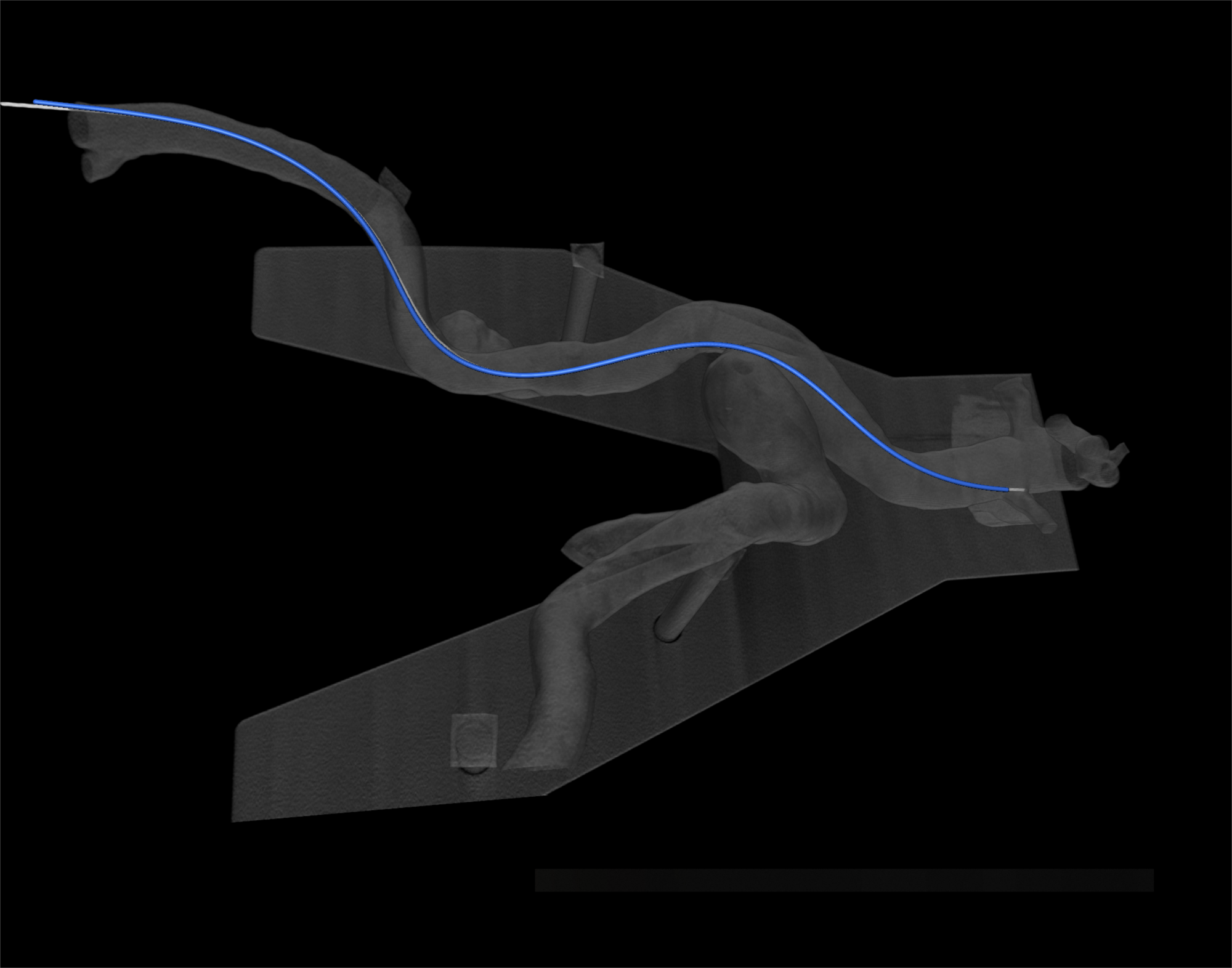}
	\end{minipage}
	\caption{The first image shows the vessel phantom with the fiber inside on a CT bed, the second image the corresponding CT scan with the reconstructed shape (blue) and ground truth (white)}	
	\label{fig-3Daorta}
\end{figure}

In the last experiment we evaluated our model in a realistic endovascular scenario and inserted our fiber into a 3D printed vessel phantom, as shown in Fig.~\ref{fig-3Daorta}. Here we obtained an average error $e_{\text{avg}}=\SI{1.13}{mm}$ and maximum error $e_{\text{max}}=\SI{2.11}{mm}$, which indicates an accurate shape reconstruction. This is also visible in the right image of Fig.~\ref{fig-3Daorta}: The reconstructed shape, represented by the blue line, fits almost perfectly to the ground truth of the CT scan.

%%%%%%%%%%%%%%%%%%%%%%%%%%%%%%%%%%%%%%%%
%%------------ CONCLUSION ------------%%
%%%%%%%%%%%%%%%%%%%%%%%%%%%%%%%%%%%%%%%%

\section{Conclusion}
In this work we presented an optimized model for shape sensing with multicore fibers for flexible instruments. We conducted a detailed error analysis for every step of the shape reconstruction procedure. The main error sources of shape sensing with multicore fibers are corrupted reference wavelengths for the wavelength shift computation, direction angles changed by the twist present in the multicore fiber and curvature values, which are distorted by using a wrong photo-elastic coefficient or wrong radii. This indicates that two calibration measurements for every FBG fiber need to be done. The first one is used to determine the Bragg wavelength $\lambda_b$ with $\varepsilon = 0$, the second one to get the twist angle $\varphi_{\text{twist}}$ where $\kappa \neq 0$. Further factors influencing the shape are the equation system defined by the used fiber configuration, the interpolation of curvature and angle values and the chosen reconstruction algorithm.

Furthermore we evaluated the accuracy of our model with 3D measurements in a CT scanner. We received an accuracy around $e_{\text{avg}} \approx$ \SIrange[range-units = single]{0.35}{1.15}{mm} and $e_{\text{max}} \approx$ \SIrange[range-units = single]{0.75}{7.53}{mm}. Finally we tested our fiber system in a real endovascular scenario and obtained high accuracies ($e_{\text{avg}}=\SI{1.13}{mm},\; e_{\text{max}}=\SI{2.11}{mm}$). These experiments show promising results for using multicore fibers for shape sensing of catheters.

In future work we aim to enable a full endovascular catheter navigation. For this purpose we plan to combine the reconstructed shape obtained by the multicore fiber with the position and orientation of a electromagnetic tracking system.

\section*{Acknowledgements} 
We thank Armin Herzog, Institute for Neuroradiology, University Hospital Schleswig-Holstein, L\"ubeck for support when using the CT scanner
and the Department of Surgery, University Hospital Schleswig-Holstein, L\"ubeck for providing the 3D vessel model printed by Fraunhofer EMB.
This work was funded by the German Federal Ministry of Education and Research (BMBF, project Nav EVAR, funding code: 13GW0228C).

\section*{Compliance with ethical standards}

\textit{Funding:} \\
This work was funded by the German Federal Ministry of Education and Research (BMBF, project Nav EVAR, funding code: 13GW0228C). \\
\\
\textit{Conflict of interest:} \\
The authors declare that they have no conflict of interest. \\
\\
\textit{Ethical approval:} \\
All procedures performed in studies involving human participants were in accordance with the ethical standards of the institutional and/or national research committee and with the 1964 Helsinki Declaration and its later amendments or comparable ethical standards. This article does not contain any studies with animals performed by any of the authors.\\
\\
\textit{Informed consent:} \\
Informed consent was obtained from all individual participants included in the study.

\bibliographystyle{spmpsci}  
\addcontentsline{toc}{chapter}{Bibliography}
% The following includes the bibliography
\bibliography{paper}
\end{document}